\documentclass[12pt]{spieman}
\usepackage{amsmath,amsfonts,amssymb}
\usepackage{graphicx}
\usepackage{setspace}
\usepackage{tocloft}
\usepackage{lineno}

\title{High-Dimensional Single-Photon Frequency Conversion for Orbital Angular Momentum Comb State}

\author[a,b]{Sheng-Qiang Jia}
\author[a,b,c]{Hao Sheng}
\author[a,b]{Zhiming Qing}
\author[d]{Li Fan}
\author[a,b]{Jianping Ding}
\author[a,b,$\ast$]{Jia-Min Xu}
\author[a,b,e]{Zhi-Cheng Ren}
\author[a,b,c,f,g,h,$\ast$]{Xi-Lin Wang}
\author[a,b,e,$\ast$]{Hui-Tian Wang}

\affil[a]{National Laboratory of Solid State Microstructures and School of Physics, Nanjing University, Nanjing 210093, China}
\affil[b]{Collaborative Innovation Center of Advanced Microstructures, Nanjing University, Nanjing 210093, China}
\affil[c]{Hefei National Laboratory, Hefei 230088, China}
\affil[d]{College of Physics Science and Technology, Institute of Applied Photonic Technology, Yangzhou University, Yangzhou 225002, China}
\affil[e]{Collaborative Innovation Center of Extreme Optics, Shanxi University, Taiyuan 030006, China}
\affil[f]{Jiangsu Physical Science Research Center, Nanjing 210093, China}
\affil[g]{Jiangsu Key Laboratory of Quantum Information Science and Technology, Nanjing University, Nanjing 210093, China}
\affil[h]{Synergetic Innovation Center of Quantum Information and Quantum Physics, University of Science and Technology of China, Hefei 230026, China}

\cftpagenumbersoff{figure}
\cftpagenumbersoff{table} 
\begin{document} 
\maketitle

\begin{abstract}
The orbital angular momentum (OAM) degree of freedom provides an important resource for constructing high-dimensional optical quantum information systems. However, the generation of high-dimensional single-photon OAM comb state and its high-fidelity frequency conversion remain challenging due to limited spatial modulation resolution, imperfect mode matching, and finite nonlinear conversion efficiency. Here, we proposed and experimentally demonstrated a scalable scheme for the generation and frequency conversion of high-dimensional single-photon OAM comb state. We prepared a single-photon OAM comb state with a dimension of up to 128, achieving a fidelity of 0.787 between the reconstructed and target states and an effective dimension of 100. Furthermore, using the sum-frequency generation, we demonstrated the frequency upconversion of single-photon high-dimensional OAM comb state from the communication band to a visible wavelength of 520~nm, while verifying the preservation of both the OAM mode distribution and phase information through density matrix reconstruction. Our work established a versatile platform for the generation, manipulation, and characterization of high-dimensional OAM comb states, with potential applications in cross-band quantum information transfer, high-capacity quantum communication, and high-dimensional quantum networks.  
\end{abstract}

\keywords{high-dimensional photonic states, orbital angular momentum comb, frequency conversion}

{\noindent \footnotesize $^\ast$Address all correspondence to Jia-Min Xu, \linkable{jiaminxu@nju.edu.cn}; Xi-Lin Wang, \linkable{xilinwang@nju.edu.cn}; Hui-Tian Wang, \linkable{htwang@nju.edu.cn}}

\begin{spacing}{2}   

\section{Introduction}

Photons are fundamental carriers of quantum information, offering low-loss propagation, weak coupling to the environment, and multiple controllable degrees of freedom.
These properties have laid a wide range of applications in quantum communication, quantum computing, and precision measurement.
Among the various degrees of freedom of photons, orbital angular momentum (OAM) provides a theoretically infinite dimensional Hilbert space and has played an important role in classical and quantum optics~\cite{willner2015optical,erhard2018twisted,erhard2020advances,fang2021nanophotonic,chen2021engineering,allen1992orbital}.
In the classical regime, orthogonal OAM modes can serve as independent channels for multiplexing, increasing the transmission capacity in optical communication \cite{wang2012terabit,bozinovic2013terabit}.
In the quantum regime, this large state space provides a natural resource for preparing high-dimensional photonic quantum states, which can substantially enhance information capacity and robustness against noise, offering new opportunities for large-scale quantum information processing \cite{groblacher2009experimental}.

Preparing coherent superpositions of OAM modes is a key step toward exploiting this high-dimensional state space for quantum information processing.
OAM combs are a class of such superpositions, comprising modes with equally spaced OAM orders and equal weights, in analogy with optical frequency combs \cite{fu2022orbital,wang2022high,liu2023frequency,ren2024ondemand,ahmed2025customizing,lu2025orbital}.
The OAM combs offer a large encoding space and substantial advantages in increasing channel capacity, expanding entanglement dimensionality, and enabling high-dimensional quantum storage \cite{groblacher2009experimental,mair2001entanglement,fickler2012quantum,dong2023highly}.
Despite these advantages, the generation and manipulation of single-photon OAM comb states become more challenging as their dimensionality increases. As OAM dimensionality increases, suppressing modal crosstalk and accurately reproducing the required spatial phase structures place stringent demands on the pixel resolution, phase modulation accuracy, and mode conversion efficiency of spatial light modulators (SLMs).
To date, OAM states spanning tens of modes have been experimentally prepared and characterized \cite{malik2014direct, erhard2020advances, Gounden2026}, but the scalable generation of high-dimensional OAM comb states is still limited by the spatial bandwidth of the optical system and the finite resolution of spatial modulators \cite{roux2003optical,pinnell2021probing}.
Beyond state preparation, the use of single-photon OAM combs across different photonic systems requires wavelength compatibility. Preparation, transmission, storage, and detection may favor different spectral ranges, requiring interfaces between photons at different wavelengths. Preservation of the encoded OAM information requires that the relative mode weights and phases be maintained under the expected mapping between input and output modes. Achieving high-fidelity wavelength conversion of high-dimensional OAM comb state is therefore a key requirement for transferring quantum information between systems operating at different wavelengths \cite{zhou2016orbital}.

Here, we propose and experimentally demonstrate a scalable scheme for generating high-dimensional single-photon OAM comb states by combining a vortex input field, a custom-designed liquid-crystal phase plate, and polarization multiplexing. Using this scheme, we prepare heralded single-photon OAM comb states with a dimension of up to 128, with the reconstructed 128-dimensional OAM state attaining a fidelity of 0.787 relative to the target state. We further demonstrate the sum-frequency conversion of the generated 4-, 8-, and 16-dimensional single-photon OAM comb states from 1560~nm to 520~nm using a 780~nm femtosecond pump laser.
A switchable optical path enables density matrix tomography of the single-photon states before and after frequency conversion, showing preservation of the modal distributions of the OAM comb states and high fidelities of the converted states relative to their target states.
Our scheme provides a scalable approach for preparing high-dimensional single-photon OAM comb states and demonstrates the frequency conversion of these states, which can support future high-capacity quantum communication and high-dimensional quantum networks.

\section{Principle and Experiment}
\label{secii}
\begin{figure*}
\includegraphics[width=1\linewidth]{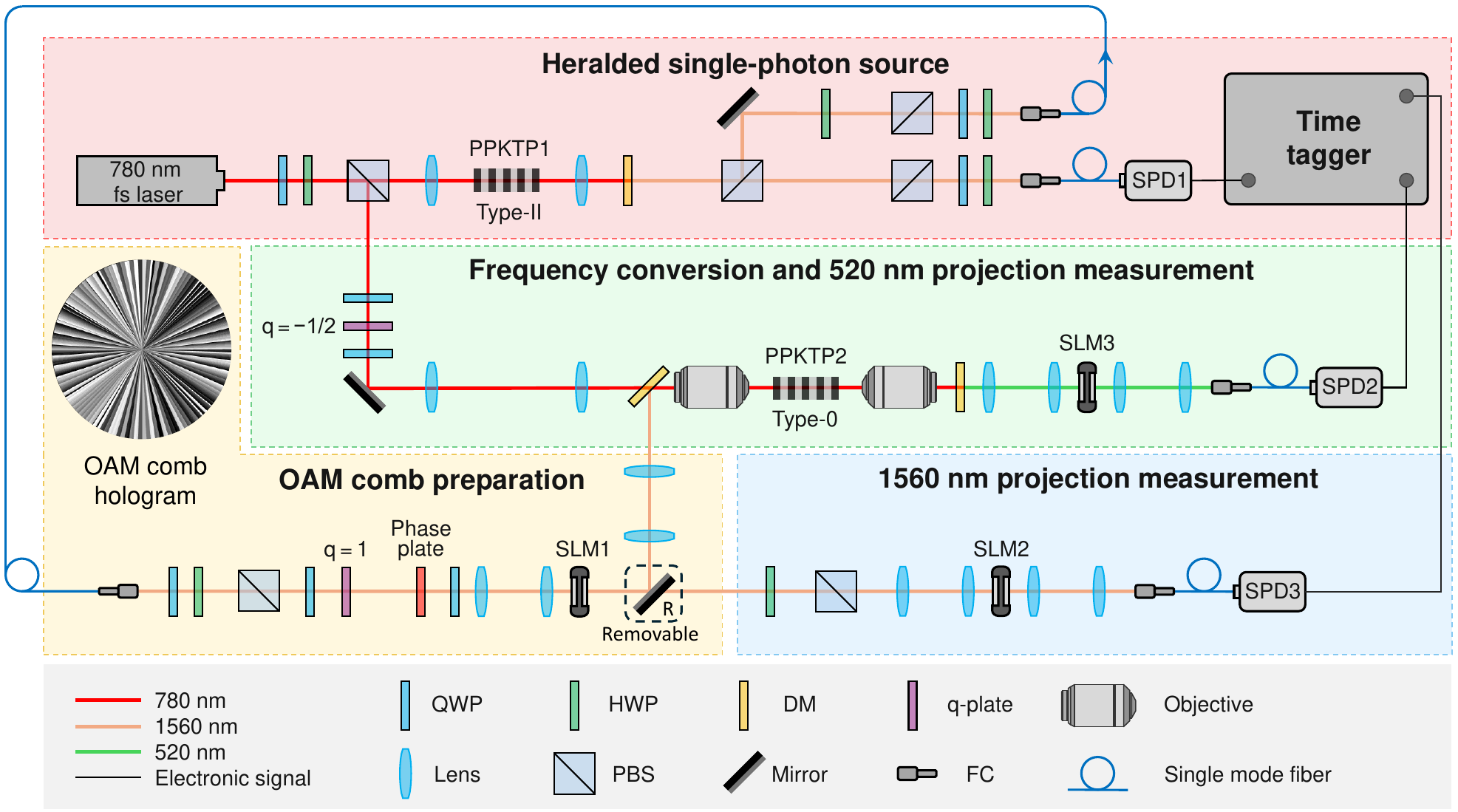}
\caption{\label{Fig1} Experimental setup for the generation, frequency conversion, and characterization of high-dimensional single-photon OAM combs. A shared 780~nm femtosecond laser pumps type-II spontaneous parametric down-conversion in PPKTP1 and type-0 sum-frequency conversion in PPKTP2. Detection at SPD1 heralds photons at 1560~nm , which are delivered through a single mode fiber to the OAM comb preparation module containing a $q$-plate (\(q=1\)), a custom liquid-crystal phase plate (the red plate), and SLM1. The OAM comb hologram is shown at the upper left, which is encoded onto the fabricated liquid-crystal phase plate. The removable mirror R selects either direct projection measurements at 1560 nm (SLM2 and SPD3) or frequency conversion followed by projection measurements at 520 nm (SLM3 and SPD2). Coincidences between SPD1 and the selected analysis detector are recorded by the time tagger. The reflective SLMs are depicted as transmissive elements for schematic clarity. QWP, quarter-wave plate; HWP, half-wave plate; PBS, polarizing beam splitter; DM, dichroic mirror; FC, fiber coupler; SLM, spatial light modulator; SPD, single-photon detector; PPKTP, periodically poled potassium titanyl phosphate.}
\end{figure*}

Figure~\ref{Fig1} shows the experimental setup, which comprises a heralded single-photon source, an OAM comb preparation module, a frequency-conversion stage, and projection measurement system. A removable mirror switches between direct characterization of the prepared states at 1560~nm and characterization after conversion to 520~nm. Because the two configurations share the source and preparation module, the input and converted states can be reconstructed separately under the same preparation conditions.

Heralded photons at 1560~nm are generated by type-II collinear spontaneous parametric down-conversion (SPDC) in a 2-mm-long periodically poled potassium titanyl phosphate (PPKTP) crystal pumped by a femtosecond laser at 780~nm. The generated photon pairs have orthogonal horizontal and vertical polarizations. Following polarization adjustment and spectral filtering, one photon is coupled into an optical fiber and directed to a single-photon detector, then heralding its partner. The heralded photon is delivered through a fiber and collimated before entering the OAM comb preparation module.

The OAM comb preparation module is based on spatial phase modulation, with the required phase patterns constructed using the two-fold symmetric multi-spiral-phase hologram design introduced in our previous work \cite{ren2024ondemand}. To generate a \(d\)-dimensional comb state, the hologram divides the azimuth into \(d\) pairs of diametrically opposed sectors of equal angular width, assigning different spiral phases and selecting their relative phase offsets to equalize the target mode weights. The resulting OAM comb hologram is shown in the inset within the yellow box in Fig.~\ref{Fig1}.

For low-dimensional comb states, the corresponding phase patterns are displayed directly on an SLM. The SLMs used throughout the experiment are reflective, but are depicted as transmissive element in Fig.~\ref{Fig1} for clarity. As the dimensionality increases, this design requires more numerous and narrower sectors, whose phase profiles become increasingly difficult to resolve near the hologram center owing to the finite pixel size of an SLM. To extend the accessible dimensionality, we propose a strategy that combines an annular input field, a custom-designed liquid-crystal phase plate, and polarization multiplexing. For a fixed sector opening angle, the transverse width of each sector increases with radius, allowing its phase profile to be sampled by more pixels at larger radii. We therefore use an annular input field to avoid illuminating the poorly resolved central region of the hologram. The phase pattern obtained using the two-fold symmetric multi-spiral-phase hologram design is encoded directly into a custom liquid-crystal phase plate during fabrication, replacing the SLM-based pattern display used for low-dimensional combs. The phase plate further improves spatial sampling through its smaller pixel size of \(8\,\text{\textmu m}\), compared with \(12.5\,\text{\textmu m}\) for the SLM.

Experimentally, in this high-dimensional configuration, a quarter-wave plate (QWP) oriented at \(45^\circ\), followed by a $q$-plate (\(q=1\)), transforms the incident horizontally polarized Gaussian field into the state \(|L,2\rangle\), corresponding to a left-circularly polarized annular vortex. Then, the custom liquid-crystal phase plate generates an equal-weight comb of \(d\) modes separated by \(\Delta\ell=2\) and reverses the polarization chirality. A subsequent QWP oriented at \(0^\circ\) converts this field into an equal superposition of horizontal and vertical polarization components carrying the same OAM spectrum. Because the SLM modulates only the horizontally polarized component, a helical phase with \(\ell=-1\) shifts the OAM orders of that component by one unit while leaving the vertically polarized component unchanged. The interleaved mode sets are then projected onto a common horizontal polarization by a half-wave plate (HWP) oriented at \(22.5^\circ\), followed by a polarizing beam splitter (PBS), producing a comb of \(2d\) consecutive OAM modes with \(\Delta\ell=1\). Thus, the phase plate needs to generate only \(d\) modes to prepare a final \(2d\)-dimensional comb state. Neglecting leakage outside the target OAM comb, the ideal state evolution is
\begin{equation}
\label{eq:comb-generation}
\begin{aligned}
|H,0\rangle
&\xrightarrow{\mathrm{QWP}@45^\circ} |R,0\rangle \xrightarrow[q=1]{q\text{-plate}} |L,2\rangle \\
&\xrightarrow{\text{phase plate}}
|R\rangle\otimes
\frac{1}{\sqrt{d}}\sum_{n=0}^{d-1}e^{i\varphi_n}|2n+3-d\rangle \\
&\xrightarrow{\mathrm{QWP}@0^\circ}
\frac{|H\rangle+|V\rangle}{\sqrt{2d}}\otimes
\sum_{n=0}^{d-1}e^{i\varphi_n}|2n+3-d\rangle \\
&\xrightarrow[\ell=-1]{\mathrm{SLM}}
\frac{|H\rangle}{\sqrt{2d}}\otimes
\sum_{n=0}^{d-1}e^{i\varphi_n}|2n+2-d\rangle \\
&\quad\quad+\frac{|V\rangle}{\sqrt{2d}}\otimes
\sum_{n=0}^{d-1}e^{i\varphi_n}|2n+3-d\rangle \\
&\xrightarrow[\mathrm{PBS}]{\mathrm{HWP}@22.5^\circ}
\frac{|H\rangle}{\sqrt{2d}}\otimes
\sum_{\ell=2-d}^{d+1}
e^{i\varphi_{\lfloor(\ell+d-2)/2\rfloor}}|\ell\rangle .
\end{aligned}
\end{equation}

Here, \(H\), \(V\), \(R\), and \(L\) denote horizontal, vertical, right-circular, and left-circular polarizations, respectively, and \(|\ell\rangle\) denotes the order of an OAM mode in the OAM comb state. The modal phases \(\varphi_n\) are determined by the designed hologram. In the final sum, \(\lfloor\cdot\rfloor\) denotes rounding down to the nearest integer, so the adjacent modes \(|2n+2-d\rangle\) and \(|2n+3-d\rangle\) share the same phase \(\varphi_n\). The parameter \(q\) denotes the structural charge of the $q$-plate. The final state in Eq.~\eqref{eq:comb-generation} is normalized after selecting the $H$-polarized PBS output. For the ideal state shown, this polarization postselection has a probability of \(1/2\), before other optical losses. In Eq.~\eqref{eq:comb-generation}, \(d\) denotes the number of modes before interleaving, and the resulting comb has a dimension of \(2d\).

For frequency conversion, the prepared single-photon OAM comb at 1560~nm is combined with a femtosecond pump at 780~nm in a second 2-mm-long PPKTP crystal. To improve spatial overlap with the annular single-photon OAM comb at 1560~nm and enhance the sum-frequency conversion efficiency, we use an annular pump whose ring size is adjusted to match that of the signal in the crystal. This pump vortex, with an OAM order of $\ell_{\mathrm{p}}=-1$, is generated by a sandwich structure composed of two QWPs oriented at \(45^\circ\) sandwiching a $q$-plate with \(q=-1/2\). Under the type-0 quasi-phase-matching condition, the horizontally polarized pump and signal undergo sum-frequency conversion to generate horizontally polarized photons at 520~nm. In an ideal collinear interaction, the interacting OAM modes obey OAM conservation, as \(\ell_{\mathrm{out}}=\ell_{\mathrm{in}}+\ell_{\mathrm{p}}\). After spectral filtering, the converted photons are directed to the 520~nm projection measurement system.

The removable mirror R switches the optical path, enabling separate projection measurements of the single-photon OAM comb states before and after frequency conversion. When the mirror R is inserted, the photons prepared in the PPKTP1 are reflected into the upper frequency-conversion path and subsequently analyzed at 520~nm; when the mirror R is removed, they propagate directly to the lower 1560~nm projection measurement system.

Both systems employ the same projection measurement scheme at their respective wavelengths. The projection states are encoded in holograms displayed sequentially on the SLM, and the projected mode is coupled into a single-mode fiber for single-photon detection. We use three families of projection states~\cite{dong2023highly} as
\begin{equation}
\label{eq:tomography-projectors}
    \begin{aligned}
        \{|\ell_m\rangle\} \quad \quad \quad (1 \leq m \leq d) \\
        \Big\{\tfrac{1}{\sqrt{2}}(|\ell_m\rangle + |\ell_n\rangle) \Big\} \quad (1 \leq m < n \leq d) \\
        \Big\{\tfrac{1}{\sqrt{2}}(|\ell_m\rangle + i|\ell_n\rangle) \Big\} \quad (1 \leq m < n \leq d)
    \end{aligned}
\end{equation}

In Eq.~\eqref{eq:tomography-projectors}, \(d\) is the dimension of the selected OAM subspace and \(\ell_m\) labels the order of OAM mode. Projections onto individual modes determine the modal populations, whereas projections onto paired superpositions with relative phases of 0 and \(\pi/2\) allow the relative phases between OAM modes to be determined. Together, these \(d^2\) projection states are sufficient to reconstruct the density matrix in the selected \(d\)-dimensional OAM subspace. The projection probabilities are inferred from coincidence counts between the heralding detector and the detector in the corresponding analysis path.

\section{Results}
\begin{figure}[t]
\includegraphics[width=1\linewidth]{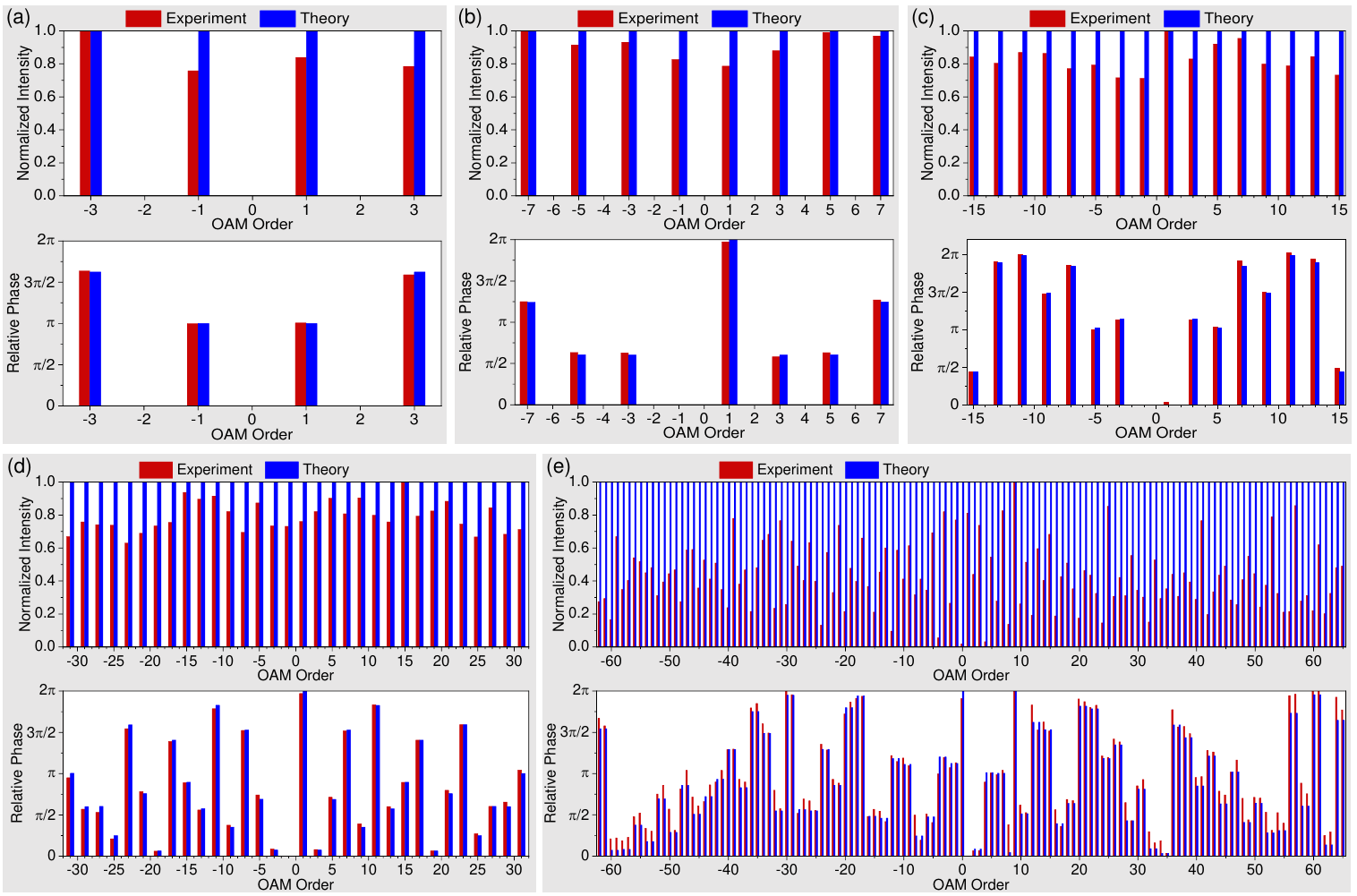}
\caption{\label{Fig2} Characterization of heralded single-photon OAM combs at 1560~nm. Upper panels aand lower panels in (a)--(e) show the modal intensity distributions and relative phases, respectively, for dimensions \(d=4,8,16,32,\) and \(128\). Red and blue bars represent reconstructed and target values, respectively. The OAM mode spacing is \(\Delta\ell=2\) for the 4-, 8-, 16-, and 32-dimensional combs and \(\Delta\ell=1\) for the 128-dimensional comb. The fidelities between the reconstructed and target states are 0.976, 0.963, 0.955, 0.929, and 0.787, respectively.}
\end{figure}

We will first prepare and characterize heralded single-photon OAM comb states with dimensions up to 128 at 1560~nm. The 4-, 8-, 16-, 32-, and 64-dimensional comb states were generated by displaying the corresponding phase patterns directly on the SLM, whereas the 128-dimensional comb was generated using the proposed high-dimensional preparation scheme described in Sec.~\ref{secii}.

For each state, coincidence counts measured with the three families of projection states in Eq.~\eqref{eq:tomography-projectors} were used to reconstruct the density matrix. As examples for five different dimensional OAM comb states, Fig.~\ref{Fig2} compare the reconstructed modal intensity distributions and relative phases with the corresponding targets. The 4-, 8-, 16-, and 32-dimensional combs have an OAM spacing of \(\Delta\ell=2\), whereas the 128-dimensional comb has \(\Delta\ell=1\) as a result of mode interleaving. Interleaving the two shifted mode sets doubles the number of modes in the target OAM comb state within a comparable OAM range. The fidelities calculated from the reconstructed density matrices quantify agreement with the target superpositions, accounting for both modal populations and intermode phases. As the target dimensions increase in order, the resulting fidelities are 0.976, 0.963, 0.955, 0.929, and 0.787, respectively.

To summarize the target mode number and the achieved state fidelity, we define an effective-dimensionality parameter \(D_{\mathrm{eff}}=dF\), where \(d\) is the target dimension and \(F\) is the fidelity relative to the target state. For the 128-dimensional state, this gives \(D_{\mathrm{eff}}\simeq100.7\). These results demonstrate that the proposed scheme supports the preparation of OAM combs at the single-photon level over a large mode space.

\begin{figure}
\includegraphics[width=0.7\linewidth]{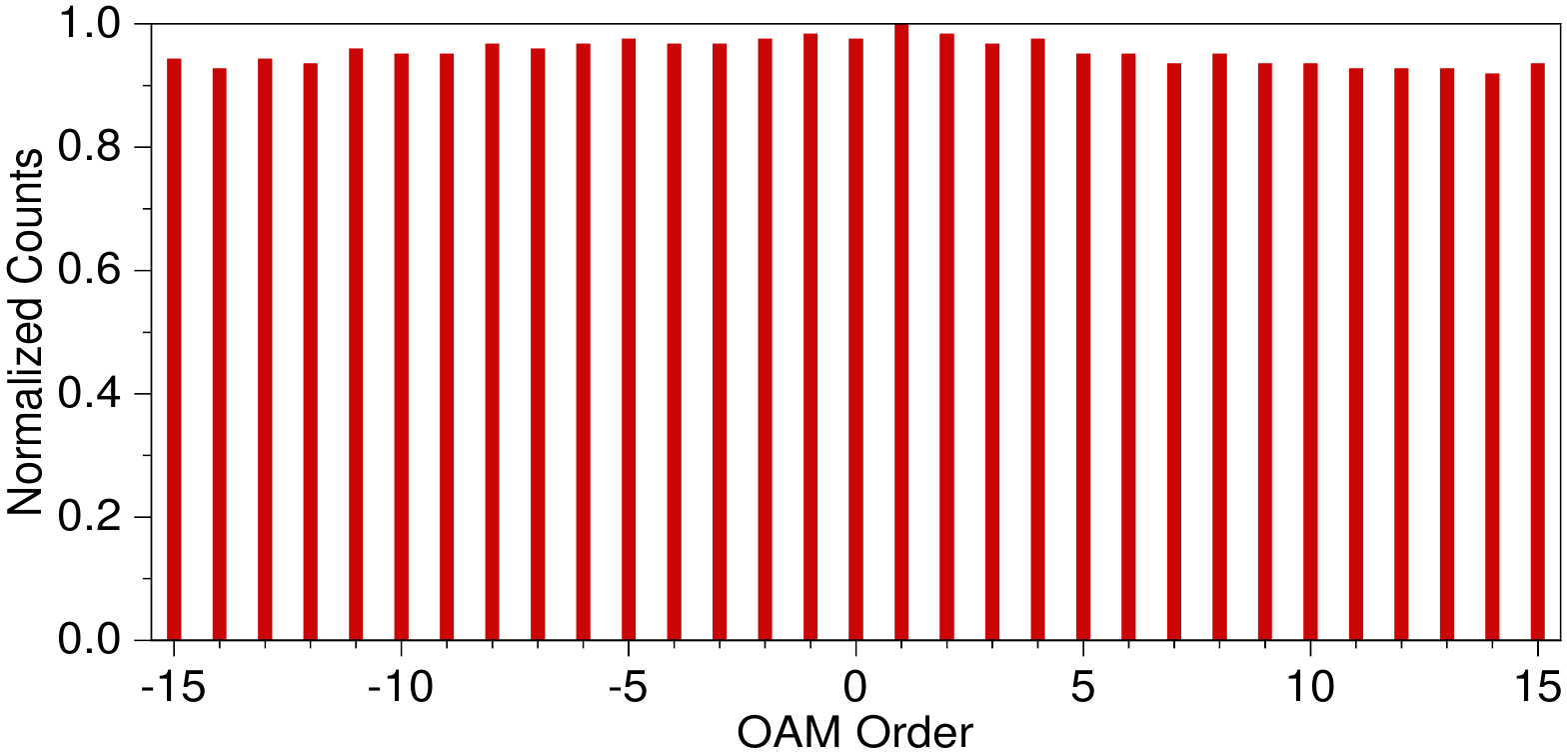}
\centering
\caption{\label{Fig3} Normalized coincidence counts for individual OAM modes after sum-frequency conversion from 1560 to 520~nm. The converted photons are projected onto the corresponding output modes using conjugate phase holograms. The counts remain relatively uniform over the tested OAM orders from \(-15\) to \(15\).}
\end{figure}

Next, we will turn to frequency conversion, beginning with measurements of the coincidence response to individual OAM modes. A $q$-plate and an SLM were used to transform an incident Gaussian field into individual OAM modes, which then carry out sum-frequency conversion with the 780~nm femtosecond pump with \(\ell_{\mathrm{p}}=-1\). The converted photons were analyzed using phase holograms conjugate with the corresponding output OAM modes. The normalized coincidence counts for the tested OAM orders from \(-15\) to \(15\) are shown in Fig.~\ref{Fig3}. These measurements show a relatively uniform combined response of the preparation, conversion, and detection stages over the tested modes, providing preliminary experimental validation for the subsequent frequency conversion of OAM combs.

We then investigated the sum-frequency conversion of the prepared single-photon OAM combs. However, efficient conversion requires precise spatial mode matching between the signal and pump, and the limited conversion efficiency in our setup results in lower coincidence counts than those obtained from direct measurements at 1560~nm. At these counts, acquiring the \(d^2\) projection measurements required for full tomography can take more than two weeks for converted combs with dimensions greater than 16. We therefore first demonstrated the frequency conversion of 4-, 8-, and 16-dimensional OAM combs from 1560 to 520~nm experimentally.

\begin{figure}
\includegraphics[width=1\linewidth]{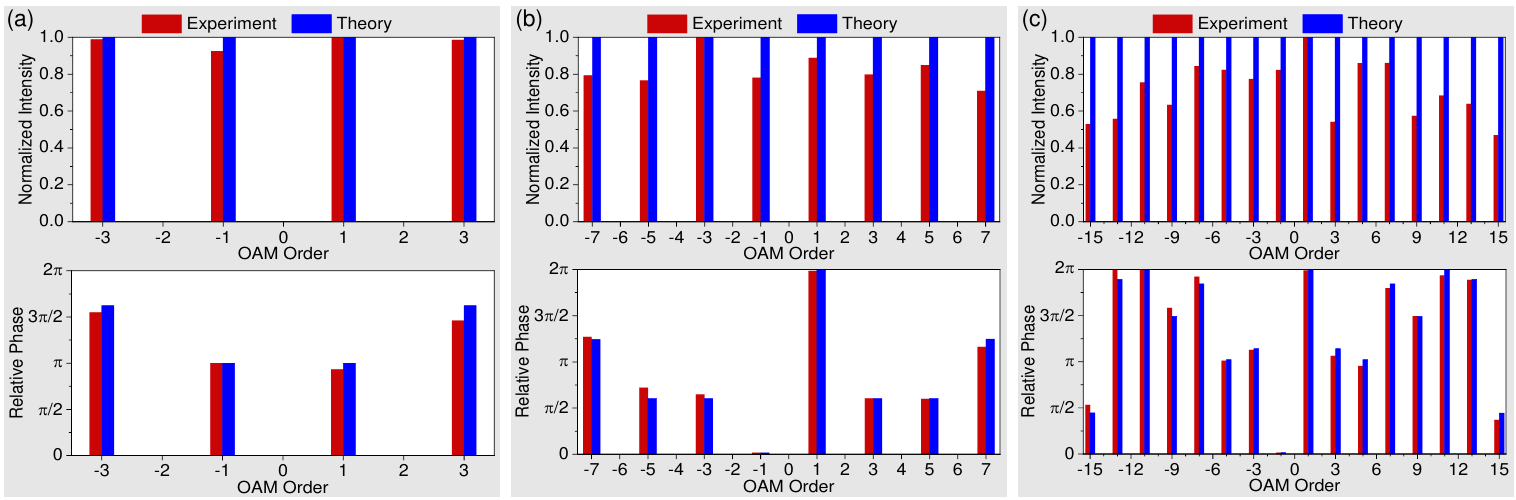}
\caption{\label{Fig4} Characterization of heralded single-photon OAM combs after sum-frequency conversion from 1560 to 520~nm. Upper panels and lower panels in (a)--(c) show the modal intensity distributions and relative phases, respectively, for dimensions \(d=4,8,\) and \(16\). Red and blue bars represent reconstructed and target values, respectively. The fidelities of the reconstructed output states relative to their targets are 0.901, 0.881, and 0.804, respectively.}
\end{figure}

Figure~\ref{Fig4} compares the reconstructed modal intensity distributions and relative phases of the converted states with the corresponding targets. From the reconstructed density matrices, we calculated the fidelities of the converted states relative to their targets, obtaining values of 0.901, 0.881, and 0.804 for the 4-, 8-, and 16-dimensional combs, respectively. Together, these fidelities and the agreement in both modal intensity and phase distributions support the transfer of the tested single-photon OAM comb states from the infrared to the visible wavelength regime.

\section{Conclusions}
We have demonstrated a scalable scheme for preparing high-dimensional OAM comb states carried by heralded single photons and manipulating them through frequency conversion. An annular pump and a custom-designed liquid-crystal phase plate improve spatial sampling, while polarization multiplexing interleaves two shifted mode sets to enlarge the accessible OAM space. The scheme enables the preparation of a 128-dimensional comb with a fidelity of 0.787 relative to its target state and an effective dimensionality of $\sim$100. We have further demonstrated sum-frequency conversion of 4-, 8-, and 16-dimensional combs from 1560 to 520~nm, with density matrix tomography supporting the preservation of modal populations and intermode phases in the tested states. The large mode space of the generated single-photon OAM combs offers a resource for high-dimensional encoding, while coherent frequency conversion allows the tested comb states to be transferred between wavelength regimes. These complementary capabilities provide an experimental basis for quantum information processing and interfaces between photonic systems operating at different wavelengths. By optimizing high-dimensional phase pattern design, improving mode-matching accuracy, and developing more efficient quantum-state measurement techniques, the fidelity and scalable dimensionality of frequency-converted high-dimensional OAM quantum states can be further enhanced.

\subsection*{Disclosures}
The authors declare no conflicts of interest.

\subsection*{Data availability}
Data underlying the results presented in this paper are not publicly available at this time but may be obtained from the authors upon reasonable request.

\acknowledgments 
This work was supported by Quantum Science and Technology-National Science and Technology Major Project (Nos. 2024ZD0300900 and 2021ZD0301400); the National Natural Science Foundation of China (Nos. 12234009, 12274215, 12404382, 12427808, 12625415, and U25A20194); the Fundamental Research Funds for the Central Universities (2026300374); the Program for Innovative Talents and Entrepreneurs in Jiangsu; Key R\&D Program of Jiangsu Province (BE2023002); the Natural Science Foundation of Jiangsu Province (Nos. BK20233001).  


\bibliography{reference}
\bibliographystyle{spiejour}   

\end{spacing}
\end{document}